TITLE

# Topological Surface Magnetism and Néel Vector Control in a Magnetoelectric Antiferromagnet

Kai Du[1], Xianghan Xu[1], Choongjae Won[2], Kefeng Wang[1], Scott A. Crooker[3], Sylvie Rangan[1], Robert Bartynski[1] and Sang-Wook Cheong[1*]

1. Department of Physics and Astronomy, Rutgers University, Piscataway, New Jersey 08854, USA
2. Laboratory for Pohang Emergent Materials and Max Planck POSTECH Center for Complex Phase Materials, Department of Physics, Pohang University of Science and Technology, Pohang 37673, Korea
3. National High Magnetic Field Laboratory, Los Alamos National Lab, Los Alamos, New Mexico 87545, USA

* To whom the correspondence should be addressed. (E-mail: sangc@physics.rutgers.edu)

**ABSTRACT**

**Antiferromagnetic states with no stray magnetic fields can enable high-density ultra-fast spintronic technologies. However, the detection and control of antiferromagnetic Néel vectors remain challenging. Linear magnetoelectric antiferromagnets (LMAs) may provide new pathways, but applying simultaneous electric and magnetic fields, necessary to control Néel vectors in**

**LMAs, is cumbersome and impractical for most applications. Herein, we show that Cr$_2$O$_3$, a prototypical room-temperature LMA, carries a topologically-protected surface magnetism in all surfaces, which stems from intrinsic surface electric fields due to band bending, combined with the bulk linear magnetoelectricity. Consequently, bulk Néel vectors with zero bulk magnetization can be simply tuned by magnetic fields through controlling the magnetizations associated with the surface magnetism. Our results imply that the surface magnetizations discovered in Cr$_2$O$_3$ should be also present in all LMAs.**

**INTRODUCTION**

Since their discovery by Louis Néel in 1930s, antiferromagnets have historically been considered only scientifically interesting but less useful for practical applications for a long time. Recently, the rapid advance of antiferromagnetic spintronics[1–5] has overturned the above perception and demonstrated that antiferromagnets with fast spin dynamics[6], low stray fields, and strong stability against magnetic fields can prevail over their ferromagnetic counterparts for faster, denser, and more robust memory devices. On the other hand, the insensitivity of antiferromagnets to external fields makes them invisible to most magnetic imaging techniques[7] and extremely difficult to manipulate. Most research efforts so far have focused on using global detection methods such as the anisotropic magnetoresistance[1], the exchange coupling[8], or the magnetic proximity effect[9] to sense the average antiferromagnetic order. Unfortunately, vital antiferromagnetic domain information is inaccessible due to the lack of spatially-resolved probes in those configurations, which hampers the full understanding of the Néel vector switching process. Antiferromagnetic states that can be easily manipulated and studied by direct magnetic imaging techniques are needed to gain insights into the Néel vector switching at the microscopic level, which are crucial for future applications of antiferromagnetic spintronic devices.

Linear magnetoelectric antiferromagnets (LMAs) have natural responses to electric and magnetic fields, which offer great potential to overcome the invisibility and stubbornness of antiferromagnetic states. As the first known and the only LMA with a Néel temperature ($T_N \approx 307$ K) above room temperature[10], $Cr_2O_3$ is an appealing candidate for room temperature antiferromagnetic devices. In particular, an intriguing net magnetization along the hexagonal $c$ axis coupled to the antiferromagnetic state has been reported in $Cr_2O_3$, which can be detected and also switched by the combination of electric and magnetic fields[8,11]. Although this net magnetization has been theoretically proposed to be an equilibrium surface magnetization[12], it is still controversial experimentally whether it is an intrinsic surface-induced magnetization[13], an extrinsic misfit-induced magnetization[14], or even a volumetric magnetization in $Cr_2O_3$ thin films[15]. The role of second phases with net magnetic moments and the dirty surface magnetism can also be relevant issues. Therefore, the microscopic mechanisms of these net magnetizations and how they respond to external magnetic fields remain elusive. Although limited magnetic imaging studies of $Co/Cr_2O_3$ heterostructures have been reported[16,17], spatially-resolved direct magnetic imaging probe of this net magnetization in single-crystal $Cr_2O_3$ is still lacking, and is crucial to solve this puzzle.

$Cr_2O_3$ forms the corundum structure consisting of a hexagonal close-packed array of oxygen anions with 2/3 of the octahedral holes occupied by chromium. Below $T_N$, spins of $Cr^{3+}$ order antiferromagnetically and collinearly along the hexagonal $c$ axis with two degenerate Néel vectors $+L$ (up-down-up-down) or $–L$ (down-up-down-up). See detailed structure in Supplementary Figure 1a. Intriguingly, non-zero diagonal linear magnetoelectric coefficients also develop both along the hexagonal $c$ axis ($\alpha_{cc}$) and in the $ab$ plane ($\alpha_{aa}=\alpha_{bb}$) below $T_N$, the signs of which are determined by its antiferromagnetic configuration[18]. For convenience, we assign $\alpha_{cc}>0$, $\alpha_{aa}=\alpha_{bb}<0$ for $+L$ domains and $\alpha_{cc}<0$, $\alpha_{aa}=\alpha_{bb}>0$ for $–L$ domains hereafter. This assignment is consistent with the opposite signs of linear magnetoelectric coefficients along the $c$- and $ab$-plane directions. Considering that the net magnetization ($Ms$) is also strongly coupled to Néel vectors[18], we here propose that it is the intrinsic surface magnetism originated from the linear magnetoelectric effect induced by the effective electric field ($Es$) at the parity-symmetry-broken surface (or interface), illustrated in Fig. 1a. Their relationship is defined by $Ms=\alpha \bullet Es$, where $\alpha$ is the linear magnetoelectric coefficient along the direction perpendicular to the surface. A natural consequence of this hypothesis is that a finite $Ms$ should exist at any surface having a non-zero $\alpha$ along the direction perpendicular to the surface. Meanwhile, the sign and strength of $Ms$ should match those of $\alpha$ for a given surface. In other words, these net magnetizations are intrinsic and topologically-protected surface magnetism. On the other hand, $Ms$ will have no relation to the magnetoelectric tensor if it originates from an extrinsic or volumetric uncompensated magnetization. Fortunately, both scenarios can be explicitly tested in single-crystal $Cr_2O_3$ if $Ms$ on both the $ab$ plane and the side surface can be directly imaged and correlated. In this work, we comprehensively examined $Cr_2O_3$ single crystals by a magnetic circular dichroism (MCD) imaging technique and show that these net magnetizations are, indeed, topological surface magnetism that exists in all crystalline orientations and is governed by its corresponding magnetoelectric coefficients at the room temperature. Consequently, a novel Néel vector switching by magnetic fields is also realized.

**RESULTS**

**Surface magnetism and surface band bending**

High-quality bulk single crystals of $Cr_2O_3$ are grown using a floating-zone (FZ) method (see methods). The crystal shows a clear antiferromagnetic transition at 307 K in the magnetic susceptibility (Fig. 1b and Supplementary Figure 1b), consistent with previous reports[19]. As expected, bulk crystals are free from net remnant magnetizations shown by the linear magnetization curve as a function of applied magnetic field (Supplementary Figure 1c and 1d). On the other hand, the polar reflective MCD sensitive to the out-of-plane magnetization shows clear signals below $T_N$ on the *ab* plane of a FZ $Cr_2O_3$ crystal (Fig. 1c), manifesting the existence of a surface magnetization coupled to the antiferromagnetic spin order. Two different spots with opposite MCD polarity can be easily found, where spot A and spot B represent antiferromagnetic domains with -*L* and +*L*, respectively. To map out the antiferromagnetic domain structures, a MCD image showing magnetization domain contrasts (+*Ms* and -*Ms*) was obtained by scanning the sample surface at 280 K (Fig. 1d), and it resembles the reported antiferromagnetic domain pattern observed by second harmonic generation[20]. The domain pattern persists until $T_N$ is reached while the domain contrast gradually fades away, leaving only background signals above 310 K (Fig. 1e-1i). Finally, a different domain pattern reappears when it is cooled back to 277 K under the zero magnetic field cooling (ZFC) condition (Fig. 1j). These results demonstrate that spatially-resolved MCD images are capable of directly capturing small net magnetizations and their underlying antiferromagnetic domain configurations in LMAs. We have also confirmed the coupling between *Ms* and Neel vectors (*L*)/magnetoelectric domains (*α*) in bulk $Cr_2O_3$ definitively by magnetoelectric field cooling experiments shown in Supplementary Figure 2.

In addition to the surface magnetism, we next show the existence of surface band bending and surface electric field in $Cr_2O_3$. It is well established that internal electric fields can develop from the band bending at the surface of semiconductors, which is widely probed by angular-resolved X-ray photoemission spectroscopy[21] (AR-XPS). Comprehensive AR-XPS studies of $Cr_2O_3$ reveal typical spectra (Fig. 1k and Supplementary Figure 3) consistent with previous XPS reports[22,23]. Moreover, we found the $Cr_2O_3$ valence band maximum bends down by ~0.2 eV at the surface on both the *ab* plane (Fig. 1k) and the side surface (Supplementary Figure 3d). Consistently, other core-level absorption peaks also are shifted by a similar amount at the surface (Supplementary Figure 3). Note that these XPS spectra are reproducible when switching between normal and grazing detection angles (Fig. 1k), suggesting that these band shifts are intrinsic and not from X-ray induced damages. Therefore, these AR-XPS results unambiguously demonstrate the existence of surface electric fields pointing into the bulk of $Cr_2O_3$ (Fig. 1l) along both orientations with a similar magnitude.

Given the above evidence, we further examine surface magnetizations on the front and the back side of the same crystal where *Es* is in the opposite direction (Fig. 2a-2d). For a thicker sample in Fig. 2a and 2b, large domains in the lower part of the image almost remain intact although some small domains in the upper part have changed their pattern. Better correlations are evident in a thinner sample in Fig. 2c and 2d, where less domain evolution is expected through the sample thickness. The same MCD polarity of these domains on the flipped surface indicates that we, indeed, have opposite surface magnetizations on the opposite surface.

**Topological surface magnetism and Néel vector controls**

Remarkably, we found these surface magnetizations are sensitive to the magnetic field when cooling through $T_N$. The sign of the MCD signal at a fixed spot can be consistently switched after a FC process (Fig. 2e) with opposite magnetic fields (± 0.5 T) applied along the *c* axis. This suggests that an unconventional Néel vector switching can be realized by magnetic fields alone, whereas the simultaneous application of both magnetic and electric fields is traditionally required for LMAs. We systematically investigated antiferromagnetic domain patterns at 280 K after a series of ZFC and FC cycles cooled down from 330 K. Random domain patterns with a domain ratio close to 50:50 are always observed after ZFC cycles on both sides of the *c*-orientated crystal (Fig. 2f and 2k). On the other hand, a positive-dominant (+*Ms*) domain pattern is observed after a +0.5 T FC cycle (Fig. 2g and 2l). After a following -0.5 T FC cycle, domain contrasts have been clearly inverted while domain patterns are almost intact (Fig. 2h and 2m), consistent with the result in Fig. 2e. Eventually, a different random domain pattern reappears after another ZFC cycle (Fig. 2l and 2n), indicating that any defect pinning effect of antiferromagnetic domain walls is negligible. Similar results have also been reproduced in another FZ and flux-grown $Cr_2O_3$ crystals (Supplementary Figure 4), demonstrating that it is an intrinsic and general property of $Cr_2O_3$. Taking differential images between +0.5 T and -0.5 T FC cycles (Fig. 2j and 2o), one can analyze the domain switching in detail where three levels of contrasts are evident. Dominant red and blue contrasts represent regions where Néel vectors have been switched in the opposite magnetic FC process, while regions with the white color (zero differential MCD signal) are small portions that are unable to switch. The ratios of non-switched areas are calculated to be only 2.8% and 4.1% for the front and the back surface, respectively. This indicates a possible competition between the magnetic-field-induced switching mechanism and the randomness from thermal fluctuations at $T_N$. Therefore, a larger magnetic field may be needed to guarantee a 100% Néel vector switching, which is not available in our current table-top experimental setup. This novel response to magnetic fields also proves that MCD signals originate from real induced magnetizations rather than the magneto-optical nonreciprocity suggested by the previous study based on pure symmetry considerations[18].

To further test if MCD signals are from intrinsic surface magnetism induced by the linear magnetoelectric effect, we examine both the perpendicular side surface (Fig. 3a) and its adjacent *ab* plane (Fig. 3b) of a FZ $Cr_2O_3$ crystal after ZFC. While clear domain patterns are observed on both surfaces (Fig. 3c and 3d), the MCD contrast level on the side surface (~3.1, Fig. 3e) is about 9 times smaller than that on the *ab* plane (~27, Fig. 3f). Considering surface electric fields have the same sign and magnitude based on previous AR-XPS results, this MCD signal ratio is qualitatively consistent with the reported ratio (~1:11) between linear magnetoelectric coefficients $\alpha_{aa}$ ($\alpha_{bb}$) and $\alpha_{cc}$ at 280 K[24]. More importantly, the merged MCD image (Fig. 3g) shows clear correlations of antiferromagnetic domains with opposite contrasts on both surfaces, which is again consistent with opposite signs of $\alpha_{aa}$ ($\alpha_{bb}$) and $\alpha_{cc}$ near room temperature.

To confirm above domain correlations, we also opt to investigate both the *ab* plane (Fig. 3h) and its perpendicular side surface (Fig. 3i) of another flux-grown $Cr_2O_3$ crystal. This flux crystal shows a slightly reduced $T_N$ at ~290 K likely due to some oxygen off-stoichiometry. Otherwise, it shows consistent MCD signals as a function of temperature (Supplementary Figure 5a and 5b) with FZ crystals. However, small domain contrasts on the side surface of flux-grown crystal are usually buried by the background curvature (Supplementary Figure 5c and 5d), making it challenging to directly visualize *Ms* domains on the side surface. To overcome this obstacle, we utilize the differential imaging method based on the *L* switching process we just discovered. The MCD image of the *ab* plane after a -0.5 T FC process was taken first (Fig. 3j), which shows a consistent negative-dominant (-*Ms*) domain pattern. We then rotate the sample and took one MCD image on the side surface followed by another image after a +0.5 T FC process (**B**//*c*). The final differential MCD image can effectively remove the background signal and reveal its intrinsic domain pattern with doubled contrasts (Fig. 3k and Supplementary Figure 5e and 5h). This time, the MCD contrast level on the *ab* plane (~27.5, Fig. 3l) is about 12 times larger than that on the side surface (5.5/2 ~2.25, Fig. 3m), which is also close to the reported ratio (~11:1) between $\alpha_{cc}$ and $\alpha_{aa}$. Once again, the merged MCD image (Fig. 3n) explicitly confirms the opposite signs of *Ms* for the same antiferromagnetic domain along different orientations. Finally, the domain ratio on the side surface (Fig. 3k) is still close to 50:50 if magnetic fields are applied along the *c* axis during the FC process, which strongly indicates there is no domain preference in the bulk and the magnetic field will only affect the domain ratio on the surface perpendicular to the field direction. We have also observed similar domain poling and Néel vector switching behaviors on the side surface after FC processes when the magnetic field is applied along the *a* axis (Supplementary Figure 6 for details). Thus, the bulk magnetization scenario can be explicitly ruled out in single-crystal $Cr_2O_3$, where a poled unequal domain ratio is also expected on the side surface when the magnetic field is applied along the *c* axis.

All these results have proved that the linear magnetoelectric contribution should be the dominating effect and responsible for the observed surface magnetism in $Cr_2O_3$ near room temperature (see Supplementary Note 1 and Supplementary Figure 7 for situations at low temperatures). More importantly, the existence of surface magnetism stemming from intrinsic surface electric fields along both principal orientations makes them topologically-protected magnetism. The magnetization of a random-cut surface will be the vector sum of the *ab* plane and the side surface components. Therefore, this surface magnetism is non-vanishing upon the reduction of specimen thickness and insensitive to the surface roughness. An intriguing implication of this conclusion is that the surface magnetizations of a single antiferromagnetic domain sample will form a three-dimensional hedgehog-type magnetization arrangement at room temperature (Fig. 4a and 4b), sharing the same topology with Néel-type magnetic skyrmions (topological charge -1 and core polarization ±1) when projected on the 2D plane. The total topological charge of such a $Cr_2O_3$ specimen will be stable against any continuous deformations of the crystal. This effective electric field ($Es$) and linear magnetoelectric mechanism can be generally applied to all LMAs with a surface spin topology depending on their non-zero linear magnetoelectric tensors. Although $Cr_2O_3$ carries only diagonal linear magnetoelectric effects, this scenario will also work universally even for materials with non-zero off-diagonal linear magnetoelectric tensors, where the final surface magnetization will simply become the vector sum of both diagonal and off-diagonal components. This demonstrates that dielectric LMAs are a new class of examples hosting topological surface states as the consequence of the ubiquitous surface electric field (*Es*), in analogy to the well-established surface Rashba effect in spin-orbit-coupled metals[25] (see Supplementary Note 2).

**DISCUSSION**

This topological surface magnetism not only enables the direct magnetic detection of its underlying antiferromagnetic domains by MCD but also can provide the essential coupling between the antiferromagnetic Néel vector and external magnetic fields. With these insights, we now discuss how the unconventional Néel vector control by magnetic fields is enabled. Starting from a balanced domain distribution with relatively straight domain walls after a typical ZFC cycle (Fig. 4c), the positive magnetic field cooling will prefer the expansion of -*L* domain on the top surface and the contraction of -*L* domain on the bottom surface (Fig. 4d). Thus, domain walls will be naturally tilted to minimize the added Zeeman energy between the surface magnetization and external magnetic fields. The domain poling effect will be evident on the surface perpendicular to the magnetic field while the overall domain ratio in the bulk still remains about 50:50. In the following negative magnetic field cooling cycle, $Cr_2O_3$ surprisingly shows inverted domain contrasts with an almost same domain pattern (Fig. 4e), which strongly implies an exotic domain wall memory effect during the field cooling process. Note that extrinsic defect pinning under the magnetic field cooling condition can be ruled out as different domain patterns can be observed in repeated field cooling experiments with different initial ZFC domain patterns (Supplementary Figure 5e and 5h). Therefore, the most plausible scenario is that some short-range spin orderings at domain walls may survive in the high-temperature paramagnetic phase. If previous domain walls are memorized and an opposite magnetic field with the same magnitude is applied upon cooling, the Néel vector at the domain region will switch to align dominating domains parallel and minor domains anti-parallel to the opposite magnetic field to minimize the total Zeeman energy (Supplementary Figure 8). Since these previous domain walls have already divided the sample surface into an unequal domain ratio favored by the same magnetic field strength, new domains and domain walls are less likely to emerge in the opposite field cooling process. However, this domain wall memory effect is less effective in ZFC cycles as both antiferromagnetic domains are allowed to nucleate randomly and emerge without any preference. Newly-formed domains and domain walls can merge with previous ones and a significantly-modified domain pattern is usually observed in ZFC cycles. As a piece of evidence, domain wall rearrangements can naturally happen in ZFC cycles indicated by back and forth jumps of the MCD signals right below $T_N$ (Supplementary Figure 5a). This memory effect also indicates exotic antiferromagnetic domain wall physics in $Cr_2O_3$, the investigation of which has just been started recently[26,27]. More advanced magnetic imaging techniques in the future with atomic resolution are needed to probe these short-range ordered spins at domain walls. Nevertheless, our proposed mechanism successfully explains all experimental observations and unveils exotic physics in the Néel vector switching process which has not been observed elsewhere.

In summary, our findings demonstrate the existence of an intriguing topological surface magnetism in $Cr_2O_3$, which is expected to be universal in all LMAs. This ubiquitous surface magnetism in LMAs enables the control of antiferromagnetic Néel vectors by magnetic fields, which provides a new pathway to manipulate antiferromagnetic states in LMAs. Furthermore, our results also illustrate the possibility of having various topological surface states induced by effective surface electric fields in insulating magnets, in addition to the well-known surface Rashba effect in conductive materials or the topological surface states of topological materials.

## METHODS

**Crystal growth and preparations**

High-quality single crystals of $Cr_2O_3$ were prepared by a floating zone technique and flux method. In the floating zone growth, high-purity $Cr_2O_3$ powder (99.9% Alfa Aesar) was filled in a rubber tube and pressed into a rod under 8000 PSI hydrostatic pressure. The rod was sintered at 1400°C for 10 hours. The crystal was grown in a laser floating zone furnace at a rate of 100 mm•h$^{-1}$ in a 0.5 L•min$^{-1}$ airflow. The as-grown crystal was annealed at 1400°C for 20 hours and cooled down to room temperature at 20°C•h$^{-1}$ before measurements. In the flux growth, the high-purity $K_2Cr_2O_7$-$K_2CrO_4$-$B_2O_3$ mixture is put in the Pt crucible with a lid and covered by $Al_2O_3$ thermal mass. The mixture was heated at 1200°C for 100 hours in the Ar gas flow with a pressure of 0.3 bar, and then cooled to room temperature. The single crystal was separated mechanically and the remaining flux was washed by distilled water. Hexagonal *ab* surface and its perpendicular side surface of crystals were oriented by Laue and hand-polished with diamond lapping films. Then, crystal surfaces were finished with polishing in the colloidal silica slurry before measurement.

Magnetic properties of oriented crystals were measured in a Magnetic Property Measurement System (MPMS, Quantum Design).

**Magnetic circular dichroism (MCD)**

Scanning MCD spectroscopy was performed using 632.8 nm light from a HeNe laser. The photon energy (1.96 eV) of this wavelength is close to the absorption peak of the d-d transition[28] of $Cr^{3+}$ in the crystal field of $Cr_2O_3$. The full wavelength dependence in the visible light range can be found in a previous report[29]. Taking reported extinction coefficient ($5.8 \times 10^4$ cm$^{-1}$) at this wave length[30], the penetration depth (inverse of the extinction coefficient) of this wavelength is ~172 nm. The real MCD probing depth should be much smaller than this value since a good fraction of the light is reflected by the very top surface. The light was linearly polarized and then modulated between right- and left-circular polarizations at 40 kHz using a photoelastic modulator. The light was focused to an approximately 20-micron spot on the sample, at a small (<5 degree) angle of incidence from the sample normal. The reflected light was detected by a photodiode and the difference between the intensity of right- and left-circular polarized reflected light was measured by a lock-in amplifier. The $Cr_2O_3$ samples were mounted on a small thermoelectric cooling/heating stage, which in turn was mounted on a motor-driven XY stage. Magnetic fields were applied by bringing a neodymium-iron-boron permanent magnet close to the sample during the cooling process. For MCD measurements below 275 K, $Cr_2O_3$ samples were mounted on the copper cold finger of a small liquid-helium optical cryostat (Oxford Microstat).

**Angular-resolved X-ray photoemission spectroscopy (AR-XPS)**

AR-XPS measurements were performed on a Thermo K-alpha instrument with a 400 μm x-ray spot of 1486.7 eV photon energy. The geometry is defined as follows: electron detection normal to the surface ($\theta=0^o$) is more bulk sensitive than electron detection at a grazing angle ($\theta=60^o$).

## DATA AVAILABILITY

The data that support the findings of this study are available within the paper as well as the Supplementary Information, and are also available from the corresponding author on reasonable request.

## ACKNOWLEDGEMENTS

The work at Rutgers University was supported by the W. M. Keck Foundation. The work at Pohang University was supported by the National Research 389 Foundation of Korea funded by the Ministry of Science and ICT (grant No. 2022M3H4A1A04074153 and 2020M3H4A2084417). The work at National High Magnetic Field Laboratory was supported by National Science Foundation (NSF) DMR-1644779, the State of Florida, and the U.S. Department of Energy (DOE). S.C. acknowledges support from the Quantum Science Center. S.R. and R.B. acknowledge the Laboratory for Surface Modification facilities for access to XPS.

## AUTHOR CONTRIBUTIONS

S.–W.C. Initiated and guided the project; X.X., C.J. and K.D. prepared the samples; K.W. and K.D. measured bulk magnetic properties; K.D. and S.C. did MCD measurements; S.R., R.B. and K.D. performed AR-XPS measurements and analysis. K.D., S.C., and S.-W.C. analysed the data and wrote the paper.

**COMPETING INTERESTS**

The authors declare no competing interests.

**Figure legends**

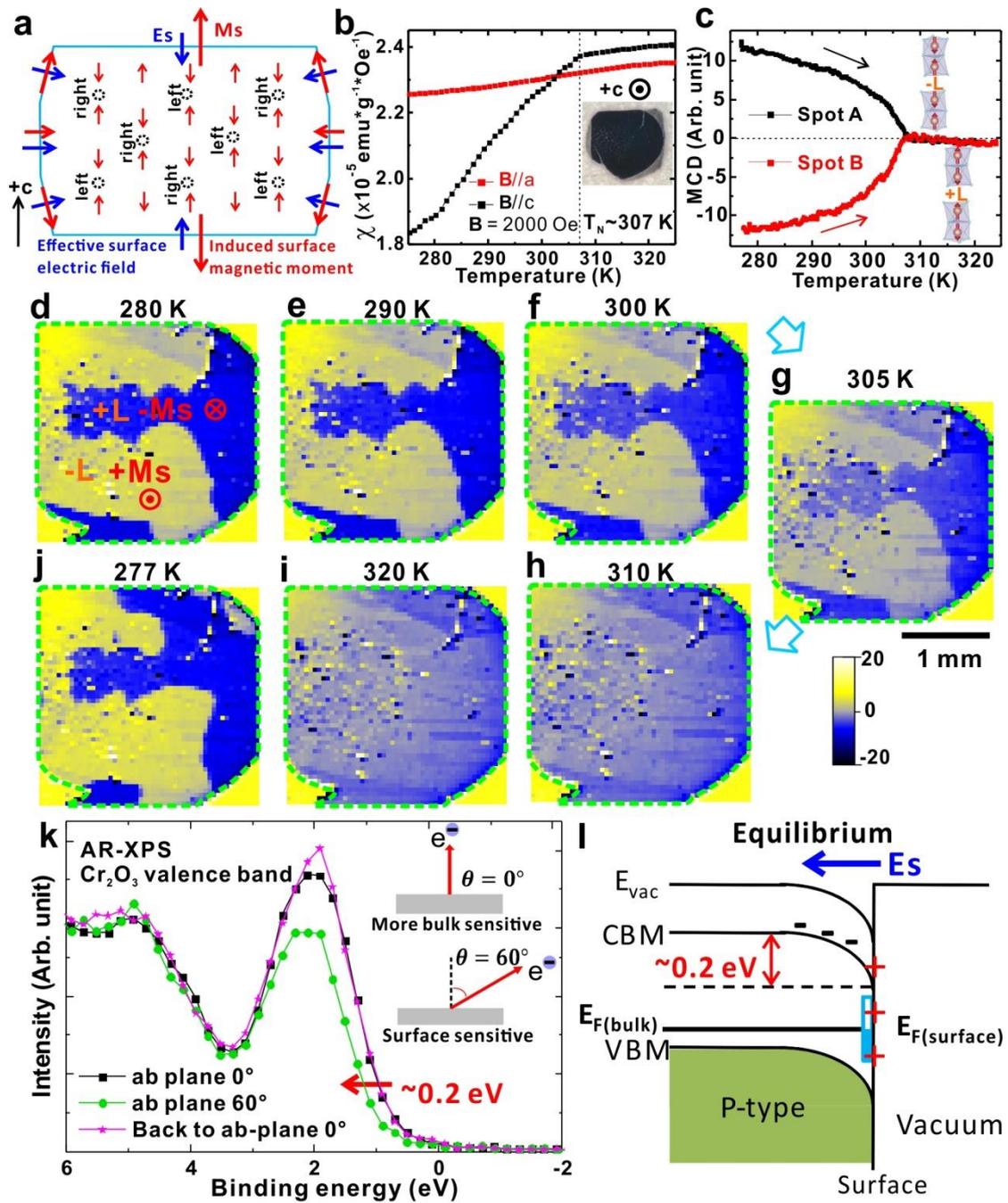

**Fig. 1: Linear magnetoelectric surface magnetism and surface electric field of Cr$_2$O$_3$.** (a) Schematics of surface magnetizations (*Ms*) induced by effective surface electric fields (*Es*). A -*L* (down-up-down-up spins) single-domain Cr$_2$O$_3$ crystal is drawn with boundaries shown by blue lines. Black dashed circles are left- or right-handed structural centers to illustrate the structural environment of spin arrangements. For an arbitrary orientation, *Es* needs to be projected onto the hexagonal *ab* plane and one of its orthogonal side surfaces. Induced *Ms* will be the vector sum of the effects in both directions. **(b)** Magnetic susceptibility as a function of the temperature of a floating-zone $Cr_2O_3$ crystal. (**B**=2000 Oe). The inset shows an optical image of the crystal. **(c)** MCD signal as a function of temperature at two fixed locations on the sample, having opposite antiferromagnetic spin structures (shown as inset) below $T_N$. **(d-j)** Sequential MCD images of a polished *ab* plane at different temperatures showing net surface magnetizations associated with antiferromagnetic domains. The domain contrasts appear below the antiferromagnetic Néel temperature $T_N \approx 307$ K. Green-dashed lines illustrate the boundaries of the crystal surface. **(k)** Angular-dependent X-ray photoemission spectroscopy of the valence band at 300 K, showing a surface band bending of ~0.2 eV. **(l)** Band diagram of p-type bulk $Cr_2O_3$ and band bending at the surface after aligning the Femi surface for equilibrium, illustrating a surface electric field pointing into the bulk.

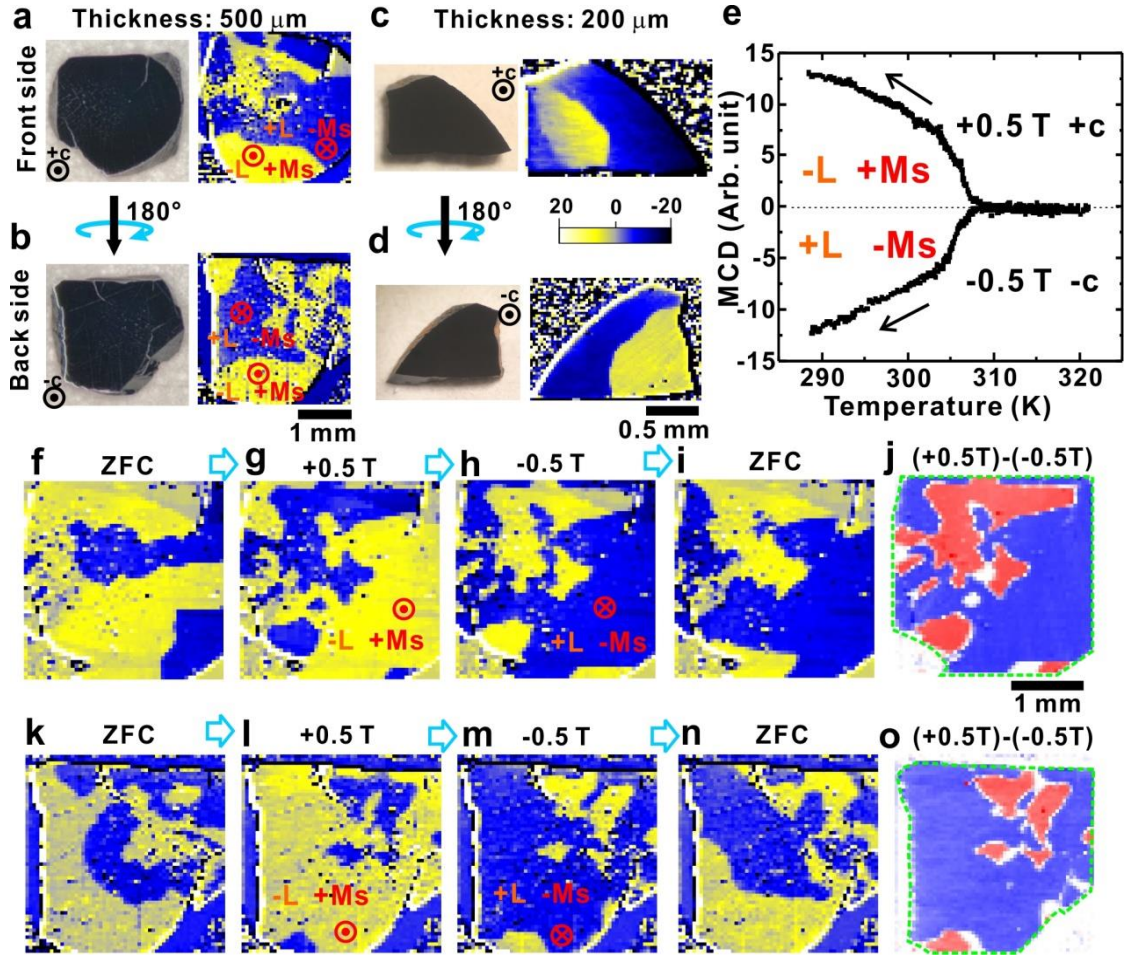

**Fig. 2: Néel vector control by cooling in opposite magnetic fields. (a)** Optical image and its MCD image at 280 K of the front *ab* plane in a 500 μm-thick crystal. **(b)** Optical image and its MCD image of the back-side *ab* plane after flipping the 500 μm-thick crystal. **(c)** Optical image and its MCD image at 280 K of the front *ab* plane in a 200 μm-thick crystal. **(d)** Optical image and its MCD image of the back-side *ab* plane after flipping the 200 μm-thick crystal. **(e)** MCD signal at a fixed spot on the sample, as a function of temperature for two opposite magnetic field (±0.5 T) cooling processes (**B**//*c*). **(f-i)** Sequential MCD images of the front surface at 280 K after cooling in different conditions from 330 K. ZFC: zero-field cooling. **B**//*c* and only applied upon cooling, which is removed before taking MCD images at 280 K. **(j)** Differential MCD image of ±0.5 T field-cooling images (g and h) showing inverted

domain contrasts. Green-dashed lines illustrate the boundaries of the crystal surface. **(k-n)** Sequential MCD images of the back surface at 280 K after cooling in different conditions from 330 K. **(o)** Differential MCD image of ±0.5 T field-cooling images (l and m) showing a similar domain inversion.

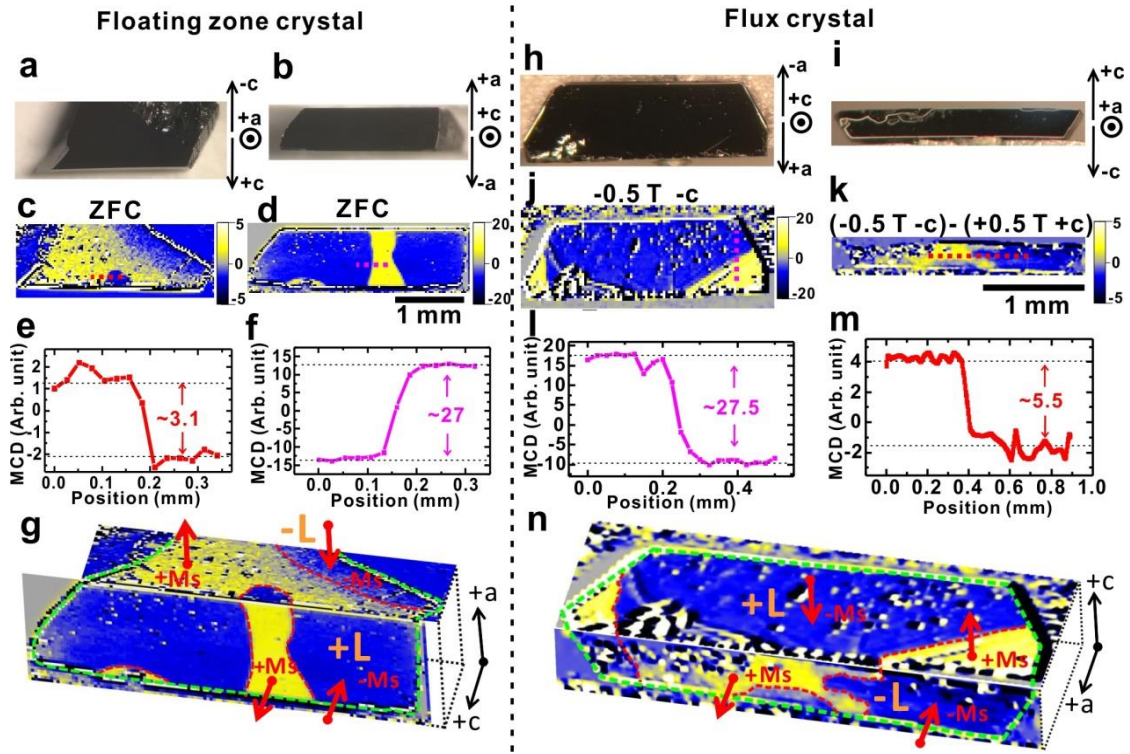

**Fig. 3: Topological surface magnetism and domain correlations. (a)** An optical image of the side surface of a FZ crystal. **(b)** An optical image of the *ab* plane of the same FZ crystal. **(c)** MCD image of the side surface at 280 K after ZFC from 330 K. **(c)** MCD image of the *ab* plane after rotating 90° at 280 K. **(e)** Line profile of the red dashed line in (c). **(f)** Line profile of the magenta dashed line in (d). **(g)** Merged MCD image on both surfaces with consistent correlations and contrasts reversals, demonstrating the topological nature of the surface magnetism. Green-dashed lines illustrate the boundaries of the crystal surface. **(h)** An optical image of the as-grown

*ab* plane of a flux-grown crystal. **(i)** An optical image of the side surface of the same flux-grown crystal. **(j)** MCD image of (h) at 280 K after -0.5 T field cooling (**B**//*c*) from 330 K. **(k)** Differential MCD image on the side surface in (i) between -0.5 T and +0.5 T field-cooling (**B**//*c*) images, representing the domain pattern after -0.5 T field-cooling process with doubled contrasts. **(l)** Line profile of the magenta dashed line in (j). **(m)** Line profile of the red dashed line in (k). **(n)** Merged MCD image on both surfaces with opposite contrasts for the same antiferomagnetic domain. Green-dashed lines illustrate the boundaries of the crystal surface. (c-d) and (j-k) share the same scale, respectively.

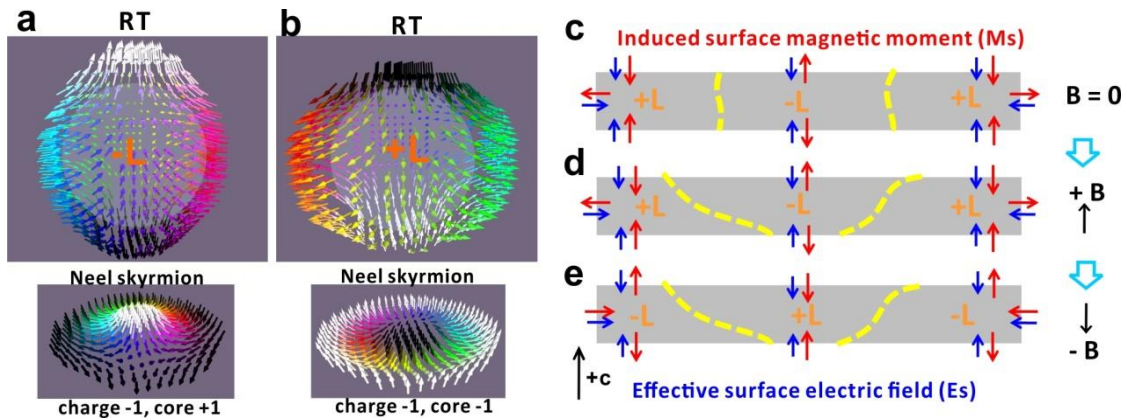

**Fig. 4: Topological surface magnetism and mechanism of Néel vector control.** **(a-b)** Schematics of surface magnetizations of a single-domain $Cr_2O_3$ crystal near room temperature (RT), and their corresponding Néel magnetic skyrmions when projected onto the plane. **(c-e)** schematics of the domain wall evolution and the Néel vector switching process in $Cr_2O_3$. Yellow dashed lines depict antiferromagnetic domain walls.

# Supplementary Information

**Title**

# Topological Surface Magnetism and Néel Vector Control in a Magnetoelectric Antiferromagnet

Kai Du[1], Xianghan Xu[1], Choongjae Won[2], Kefeng Wang[1], Scott A. Crooker[3], Sylvie Rangan[1], Robert Bartynski[1] and Sang-Wook Cheong[1*]

4. Department of Physics and Astronomy, Rutgers University, Piscataway, New Jersey 08854, USA
5. Laboratory for Pohang Emergent Materials and Max Planck POSTECH Center for Complex Phase Materials, Department of Physics, Pohang University of Science and Technology, Pohang 37673, Korea
6. National High Magnetic Field Laboratory, Los Alamos National Lab, Los Alamos, New Mexico 87545, USA

\* To whom the correspondence should be addressed. (E-mail: sangc@physics.rutgers.edu)

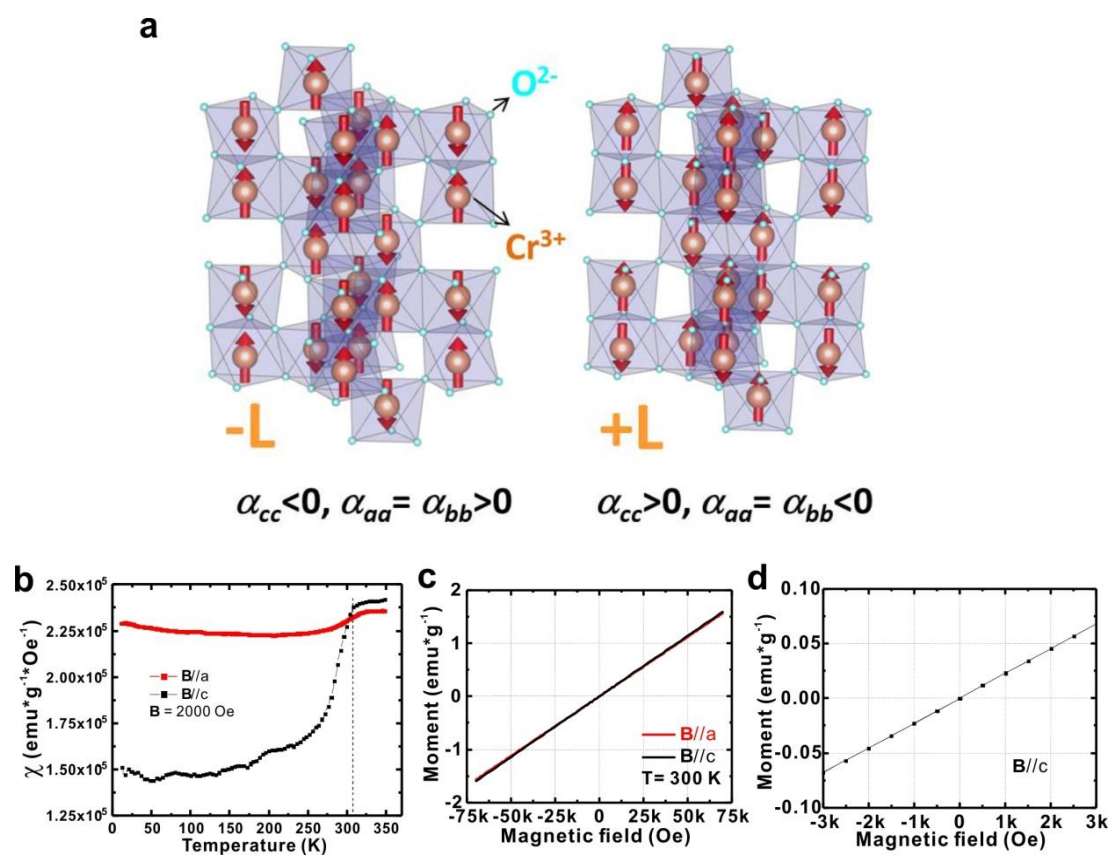

**Supplementary Figure 1: Magnetic properties of $Cr_2O_3$.** (a) Atomic and magnetic structures of two degenerate antiferromagnetic states below $T_N$. Red arrows are spins of $Cr^{3+}$. (b) Magnetic susceptibility as a function of temperature down to 5 K (**B**=2000 Oe). (c) Magnetic moment as a function of applied field for both in-plane and out-of-plane directions at 300 K, showing a perfect linear dependence without any remnant moments. (d) Zoomed center part of the out-of-plane curve in (c).

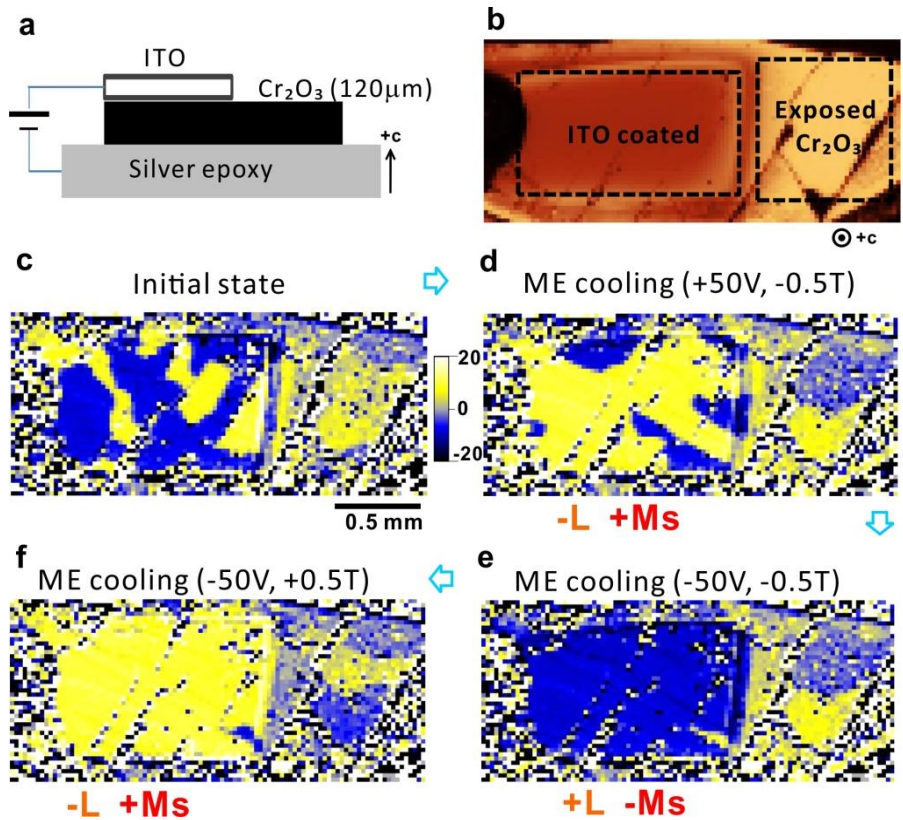

**Supplementary Figure 2: Coupling between surface magnetism and Néel vectors.** (**a**) Side-view geometry of the device for magnetoelectric (ME) field cooling experiments. A 20 nm transparent indium tin oxide film is sputtered as the top electrode. (**b**) Total reflection image of the sample showing the ITO-coated region and exposed $Cr_2O_3$ by dashed squares. (**c**) MCD image at 280 K of the initial state. (**d**) MCD image at 280 K after ME cooling (+50 V, -0.5 T) from 315 K, showing dominating +$Ms$ associated with –$L$ domain. Both electric and magnetic fields are removed before taking the image. Minor opposite domains near the edge of ITO electrodes are due to the competition between ME field (favoring –$L$, +$Ms$) and magnetic field (favoring +$L$, -$Ms$). (**e**) MCD image after ME cooling (-50 V, -0.5 T) showing fully poled +$L$ domain with -$Ms$, as ME field and magnetic field have the same preference in this configuration. (**f**) MCD image after ME cooling (-50 V, +0.5 T) showing fully poled -$L$ domain with +$Ms$,

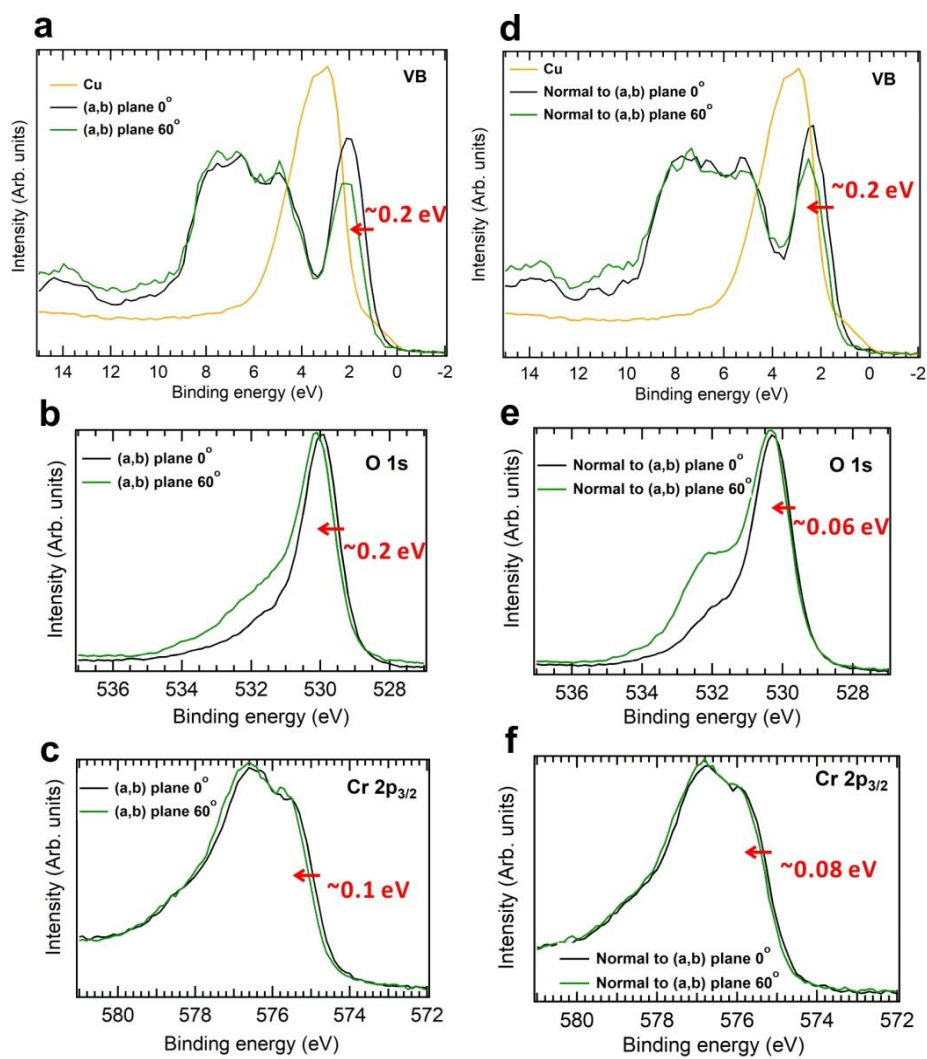

**Supplementary Figure 3: AR-XPS and surface band bending.** (a) Angular-dependent valence band spectra of the *ab* plane compared to the valence band spectrum measured on a copper foil in electrical contact with the sample. (Orientation: 0° bulk sensitive and 60° surface sensitive) (b) Angular-dependent oxygen 1s peak of the *ab* plane. (c) Angular-dependent Cr $2p_{3/2}$ peak of the *ab* plane. (d) Angular-dependent valence band spectra of the side surface (normal to the *ab* plane) and Cu foil reference valence band spectrum. (e) Angular-dependent oxygen 1s peak of the side surface. (f) Angular-dependent Cr $2p_{3/2}$ peak of the side surface.

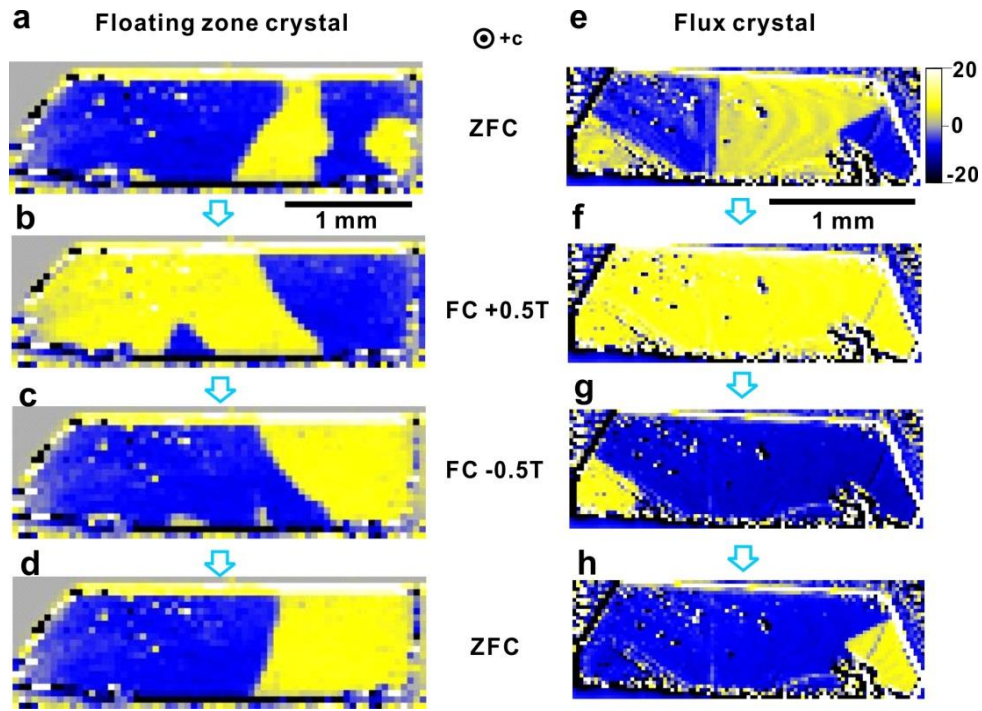

**Supplementary Figure 4: Reproducibility of Néel vector control by magnetic field cooling. (a-d)** Sequential MCD images of another FZ crystal at 280 K after cooling in different conditions from 330 K. ZFC: zero-field cooling. **B**//c and only applied upon cooling, which is removed before taking MCD images. **(e-h)** Sequential MCD images of another flux-grown crystal at 280 K after cooling in different conditions from 330 K. Domain inversion behaviors similar to Figure 2 are observed after opposite magnetic field cooling.

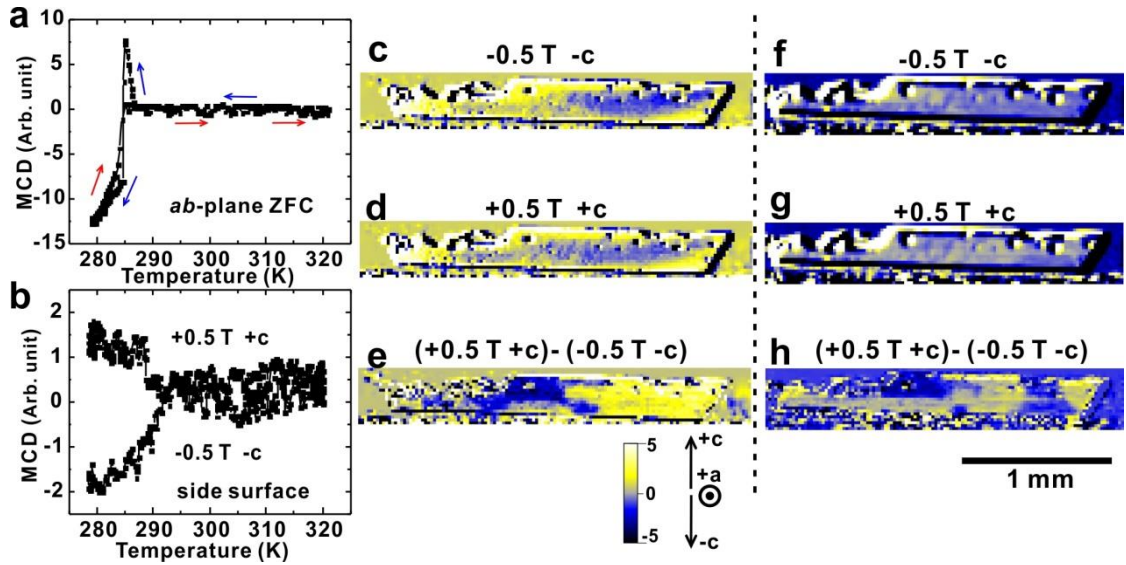

**Supplementary Figure 5: Differential MCD images on the side surface. (a)** MCD signal as a function of temperature on the *ab* plane of the flux-grown crystal in the zero-field condition. The back and forth jumps indicate domain wall movements right below $T_N$. **(b)** MCD signal as a function of temperature on the side surface of the flux-grown crystal in two opposite magnetic field (±0.5 T, **B**//*c*) cooling process. **(c)** MCD image of the side surface at 280 K after -0.5 T field cooling (**B**//c) from 330 K. The image is acquired after taking the image in Figure 3c and rotating the sample by 90°. **(d)** MCD image of the side surface after another +0.5 T field cooling (**B**//c) from 330 K. The sample was heated up to 330 K in zero magnetic field and cooled back to 280 K with a +0.5 T magnetic field. All magnetic fields are removed after reaching 280K before taking images. (e) The differential MCD image of (c) and (d), which unveils the antiferromagnetic domain pattern on the side surface. **(f-g)** Another set of MCD images of the side surface after ±0.5 T field cooling (**B**//c) from 330 K. An additional zero-field cooling cycle is added before these FC cycles to reset a random domain pattern. **(h)** The differential MCD image of (f) and (g), showing a different domain pattern on the side surface. This can rule out the extrinsic defects pinning effect in the sample.

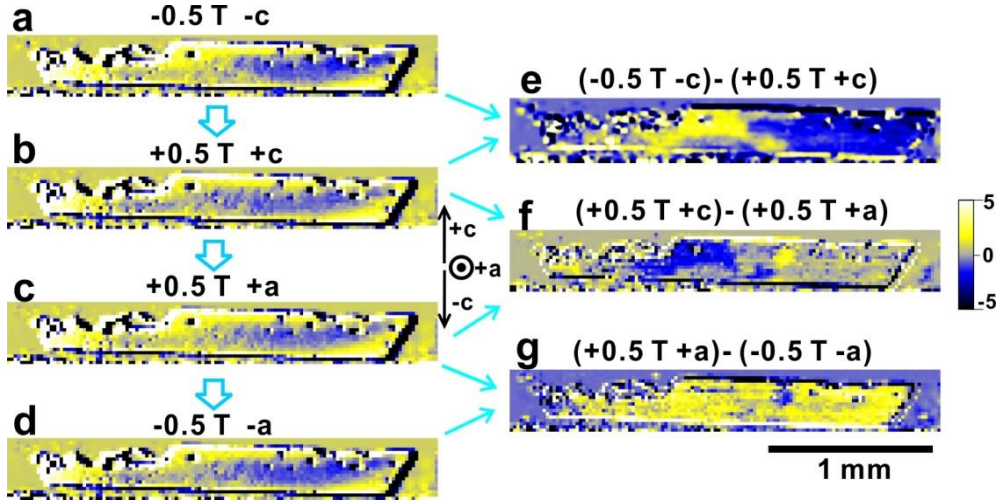

**Supplementary Figure 6: Néel vector switching on the side surface.** Sequential MCD images of the side surface at 280 K after cooling in **(a)**-0.5 T (**B**//*c*), **(b)**+0.5 T (**B**//*c*), **(c)**+0.5 T (**B**//*a*), and **(d)**-0.5 T (**B**//*a*) conditions from 330 K. **(e-f)** Corresponding differential MCD images. The existence of three contrast levels in (f) manifests a drastic domain pattern change after the 90° switching of magnetic field direction during the cooling. Meanwhile, the 180° switching of magnetic field direction during the cooling will consistently have the same domain pattern with inverted contrasts shown in (e) and (g).

**Supplementary Note 1: MCD signal contributions at low temperatures and near room temperature.**

Having established the mechanism of the topological surface magnetism near room temperature, we also checked the MCD signal at a fixed point on the *ab* plane down to 5 K (Supplementary Figure 7a). Interestingly, we found the surface magnetization does not go all the way down to zero or negative values below 85 K, as suggested by the reported temperature-dependent $\alpha_{cc}$ curve (Supplementary ref. 1). Instead, the MCD signal only drops a little before reaching a stable finite value in the low-temperature range. The non-vanishing MCD signal indicates that another independent mechanism may also induce surface magnetization on the *ab* plane, and dominate MCD signals at low temperatures when the linear magnetoelectric contribution is supposed to fade. The most probable scenario is the orphan spin mechanism that is specific to the spin structures of $Cr_2O_3$ illustrated in Supplementary Figure 7b. The up and down spins in the antiferromagnetic state can be considered as a pair of dimerized spins in either left-handed or right-handed structural sites. Since the next nearest site has a one-spin shift along the *c* axis, one kind of spin will inevitably lose its paring counterparts and act as special orphan spins when terminated at the surface. These orphan spins are not fully compensated by their surrounding dimerized opposite spins and can also contribute to the roughness-insensitive magnetization on the top and bottom surface with opposite magnetizations, especially at low temperatures when the linear magnetoelectric contribution is vanishing.

However, the dominating effect near the room temperature should still be the linear magnetoelectric mechanism due to the following reasons. First of all, this orphan spin mechanism can not explain the MCD signal on the side surface as it is specific to the *ab* plane. Therefore, the linear magnetoelectric contribution should exist. Second, the measured MCD signal ratio (9:1 for FZ crystal and 12:1 for flux crystal) between the *ab* plane and the side surface near room temperature is close to the theoretical value (~11:1) predicted by the linear magnetoelectric mechanism, which indicates very limited contributions from orphan spins near room temperature. Third, these orphan spin contributions likely scale with the magnetic moment of $Cr^{3+}$ below $T_N$, which is relatively small near room temperature and only becomes dominant at low temperatures. Therefore, we can safely conclude the linear magnetoelectric mechanism is the dominating effect at the room temperature.

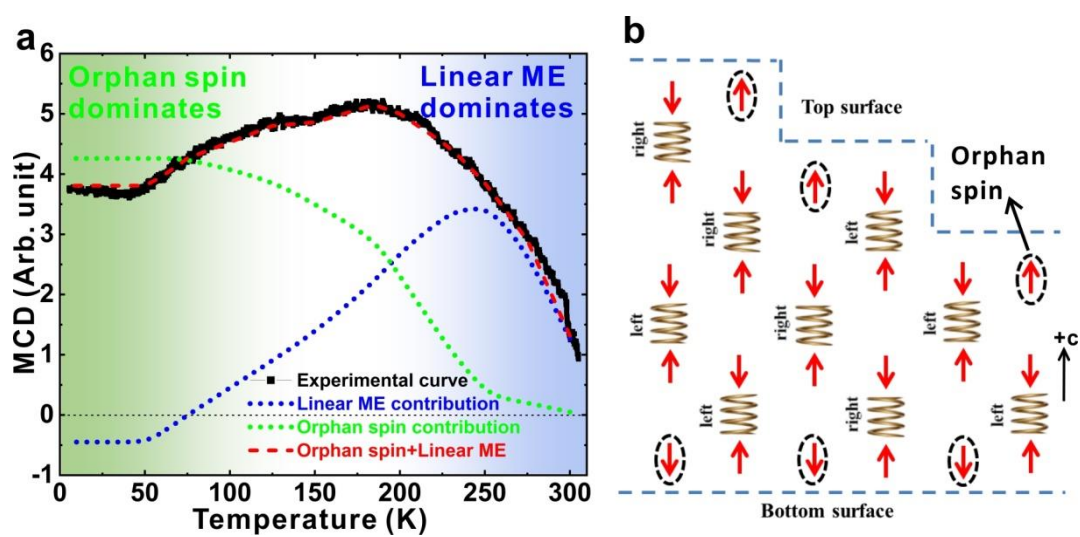

**Supplementary Figure 7: Low-temperature MCD signals and orphan spin contributions.** (**a**) Temperature-dependent MCD signals down to 5 K (black curve). The dotted curves are estimated linear magnetoelectric contributions (blue) and orphan spin contributions (green) to the MCD signal. Red dotted curve is the added total signal which overlaps with the measured MCD curve. (**b**) Schematics of surface magnetizations induced by orphan spins on the *ab* plane, which likely scales with the magnetic moment of $Cr^{3+}$ below $T_N$ and follows a curve exampled by the green dotted curve in (a).

**Supplementary Note 2: Estimating the effective surface electric field *Es***

Generally, the effective electric field (*Es*) exists ubiquitously in any symmetry-breaking surfaces or interfaces. On the other hand, it is often inert and difficult to detect as no electric fields are leaking out of the surface. However, it can enable exotic topologically-protected surface states when certain criteria are met. One prime example is the well-established surface Rashba effect in metals with spin-orbital couplings (Supplementary ref. 2). The effective electric field at the surface together with the spin-orbit coupling will induce effective magnetic fields that are locked to the momentum vector, known as spin-momentum locking. However, examples of insulating materials are still lacking so far. Dielectric linear magnetoelectric antiferromagnets (LMAs) like $Cr_2O_3$ in this work represent a new class of examples hosting topological surface states as the consequence of the effective surface electric field (*Es*).

Utilizing the measured surface magnetization and the linear magnetoelectric coefficient $\alpha_{cc}$, one can estimate the effective surface electric field value of an insulator like $Cr_2O_3$ which is otherwise challenging to quantify. Taking the reported Kerr rotation (~4 mdeg at 280 K, Supplementary ref. 3) and the typical magneto-optical strength of ferromagnetic materials like MnBi (~1.2 deg per 600 emu•cm$^{-3}$, Supplementary ref. 4), we can have the surface magnetism $Ms$ ~ 0.005 $\mu B \cdot Cr^{-1}$ at the surface for $Cr_2O_3$. Since $\alpha_{cc}$ is ~4 ps•m$^{-1}$ at 280 K (Supplementary ref. 5), we can estimate the effective surface electric field $Es = \mu_0 * Ms/\alpha_{cc} = 0.6$ V•nm$^{-1}$ for $Cr_2O_3$, where $\mu_0$ is the magnetic permeability of free space.

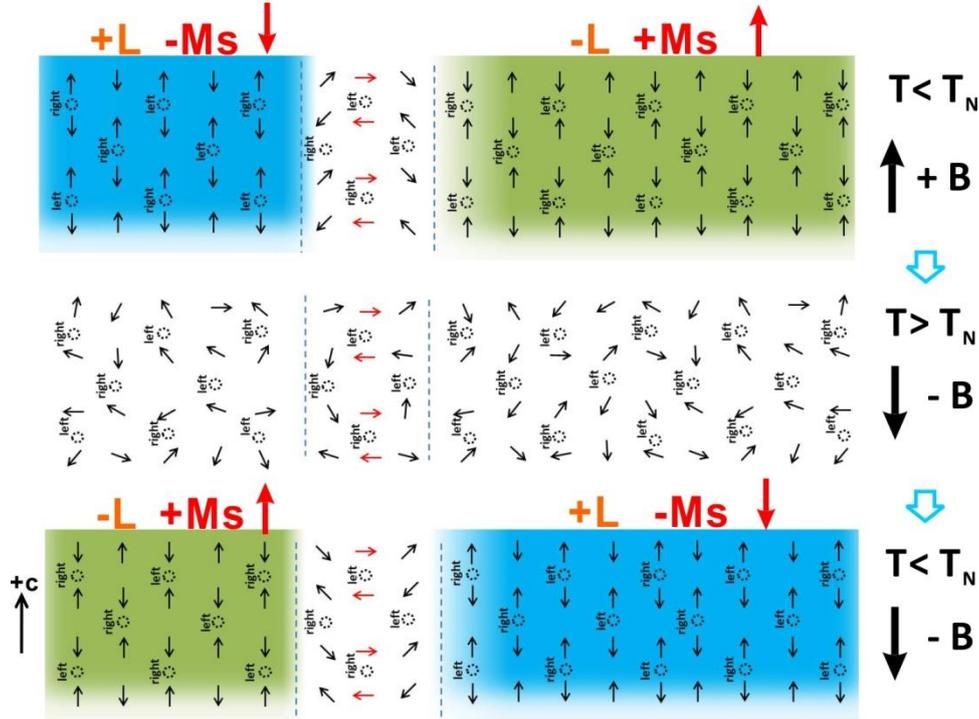

**Supplementary Figure 8: Short-range magnetic orderings at domain walls above $T_N$.** Schematics of proposed residual short-range magnetic orderings at antiferromagnetic domain walls (red arrows) in the paramagnetic phase, which enables the exotic Néel vector switching during the field cooling process in $Cr_2O_3$. Black dashed circles are left- or right-handed structural centers to illustrate the structural environment of spin arrangements. When the domain wall position is memorized above $T_N$ and magnetic field is switched, the dominating domain region (the right part of the figure) will follow the magnetic field direction and switch its Néel vector to minimize the total Zeemen energy.

**Supplementary References**